\documentclass[final]{cvpr}

\usepackage{times}
\usepackage{epsfig}
\usepackage{graphicx}
\usepackage{amsmath}
\usepackage{amssymb}
\usepackage{multirow}
\usepackage{subfigure}
\usepackage{graphicx}
\usepackage{booktabs} 
\usepackage{setspace}
\usepackage{caption}
\usepackage{float}
\usepackage{url}

\usepackage[pagebackref=true,breaklinks=true,colorlinks,bookmarks=false]{hyperref}

\begin{document}

\title{LAMP: Label Augmented Multimodal Pretraining}

\author{%
Jia Guo \\
Zhejiang University\\
{\tt\footnotesize jiuzhou@zju.edu.cn}
\and

Chen Zhu \\
  Alibaba  Group  \\
  {\tt\footnotesize none.zc@alibaba-inc.com}
\and

Yilun Zhao \\
Zhejiang University \\
{\tt\footnotesize zhaoyilun@zju.edu.cn}
\and

Heda Wang \\
  Alibaba Group \\
  {\tt \footnotesize heda.whd@alibaba-inc.com}
  \and

Yao Hu \\
  Alibaba Group \\
  {\tt\footnotesize yaoohu@alibaba-inc.com}
  \and
  
Xiaofei He \\
  Zhejiang University \\
  {\tt\footnotesize xiaofei\_h@qq.com}
  \and
Deng Cai \\
  Zhejiang University \\
  {\tt\footnotesize dengcai78@qq.com}
}


\maketitle

\begin{abstract}
Multi-modal representation learning by pretraining has become an increasing interest due to its easy-to-use and potential benefit for various Visual-and-Language~(V-L) tasks. However its requirement of large volume and high-quality vision-language pairs highly hinders its values in practice. In this paper, we proposed a novel label-augmented V-L pretraining model, named LAMP, to address this problem. Specifically, we leveraged auto-generated labels of visual objects to enrich vision-language pairs with fine-grained alignment and correspondingly designed a novel pretraining task. 
Besides, we also found such label augmentation in second-stage pretraining would further universally benefit various downstream tasks. 
To evaluate LAMP, we compared it with some state-of-the-art models on four downstream tasks. 
The quantitative results and analysis have well proven the value of labels in V-L pretraining and the effectiveness of LAMP.




\end{abstract}

\section{introduction}

Recently, due to the easy-to-use and strong representation ability of pretrained models, pretrain-then-finetune learning has been widely adapted in computer vision~(CV) and natural language processing~(NLP). This naturally draws the attention of a lot of researchers in Vision-and-Language~(V-L) field, where most of tasks, such as captioning~\cite{chen2015microsoft} and visual question answering~\cite{antol2015vqa}, also highly rely on effective cross-modal representation.

Some pioneering works, such as LXMERT~\cite{tan2019lxmert} and ViLBERT~\cite{lu2019vilbert}, have shown the potential of pretrain-then-finetune learning for V-L tasks. Most of them followed the intuition of BERT~\cite{devlin2018bert} and designed some cross-modal pretraining tasks, including aligning images with the corresponding sentences~(i.e. instance level alignment) and aligning objects in images and entities in texts~(i.e. objects-entities level alignment), to bridge the gaps between visual and language.

\begin{figure}
    \centering
    \includegraphics[width=0.95\linewidth]{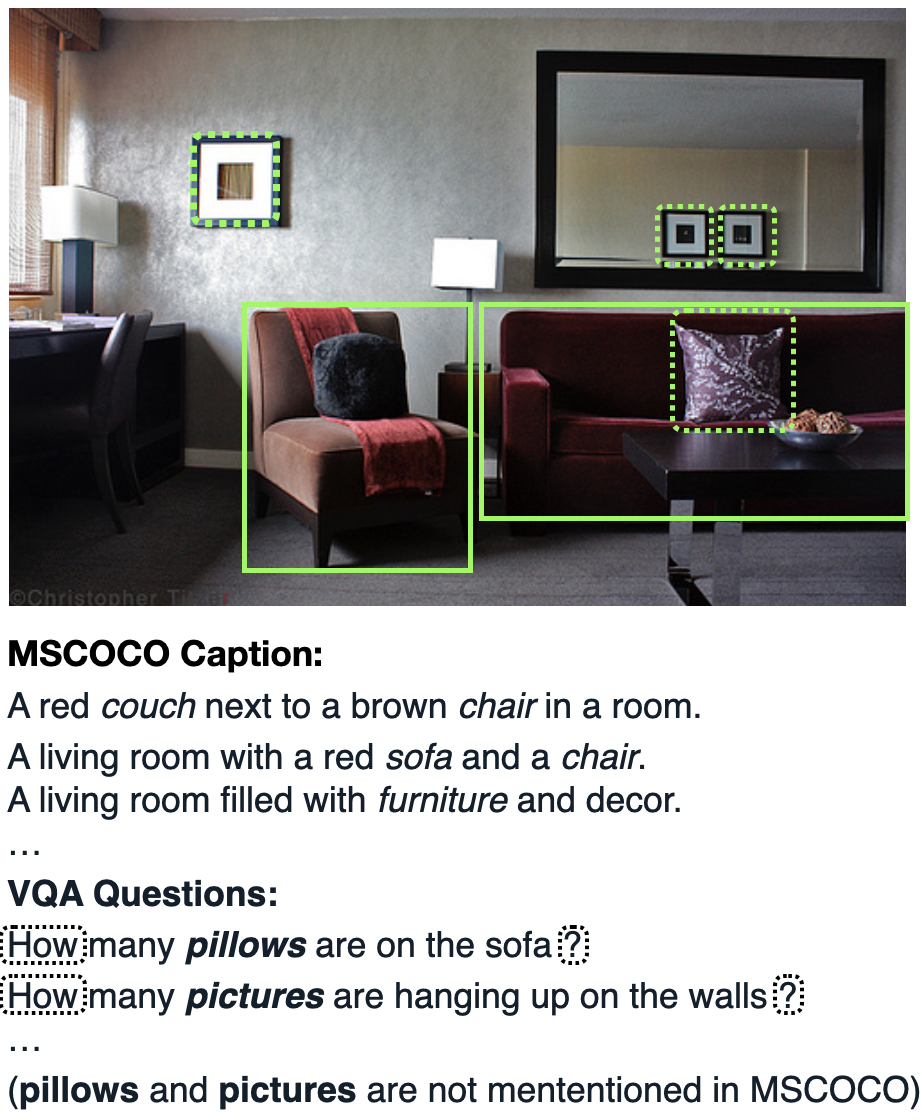}
    \vspace{-0.2cm}
    \caption{A illustration of miss-alignment problem. Solid boxes are mentioned in MSCOCO, but the dashed ones are not. Many boxes are required in VQA task while not mentioned in captions.}
    \label{fig:example}
    \vspace{-0.6cm}
\end{figure}

However, two problems arise during the adaption from BERT to multi-modal representation learning.
i. Most of current V-L pretraining datasets~(e.g. MSCOCO~\cite{chen2015microsoft} and Visual Genome~(VG)~\cite{krishna2017visual}) compose of image-caption pairs. In pretraining, captions are used to predict the objects in images. But captions often cannot provide comprehensive description for images. 
Figure~\ref{fig:example} is an example of image-caption pair. We can find the caption only cover two objects~(i.e. ``couch'', ``chair'') in the image.
The lack of textual description of other objects~(i.e. pillow) probably distorts objects-entities level alignment in pretraining.
ii. The semantic discrepancy of textual input between pretraining tasks and downstream tasks is another obvious problem in the V-L field. Also take Figure~\ref{fig:example} as an example, ``what'' and ``how'' are high-frequent words in the Visual Question Answering~(VQA) task, but both of them rarely appear in captions of images. The gaps probably weaken the representation learned by pretrained models. 

\begin{figure}[!t]
\centering
\subfigure[Example of objects, captions and labels used in LAMP.]{
\begin{minipage}[b]{0.95\linewidth}
\includegraphics[width=1\linewidth]{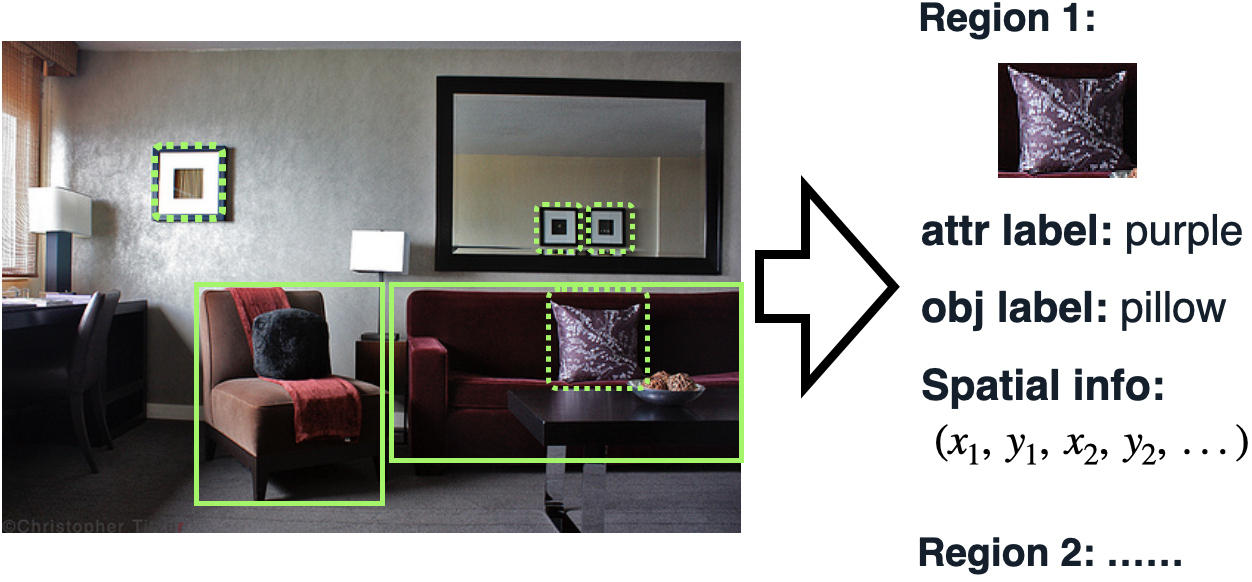}
\end{minipage}
}
\subfigure[LAMP Model for objects-labels pairs.]{
\begin{minipage}[b]{0.95\linewidth}
\includegraphics[width=1\linewidth]{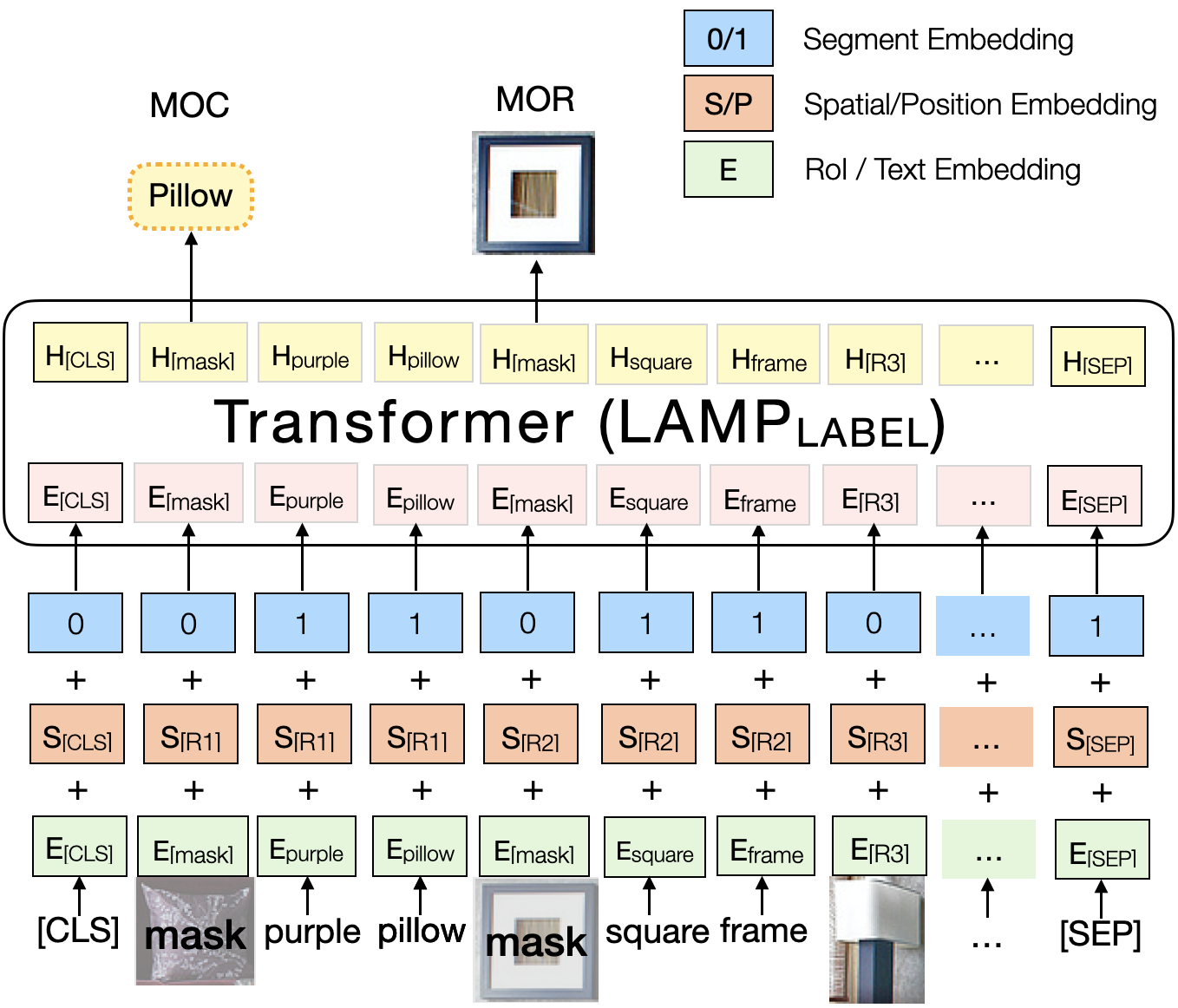}
\end{minipage}
}
\subfigure[LAMP Model for objects-caption pairs.]{
\begin{minipage}[b]{0.95\linewidth}
\includegraphics[width=1\linewidth]{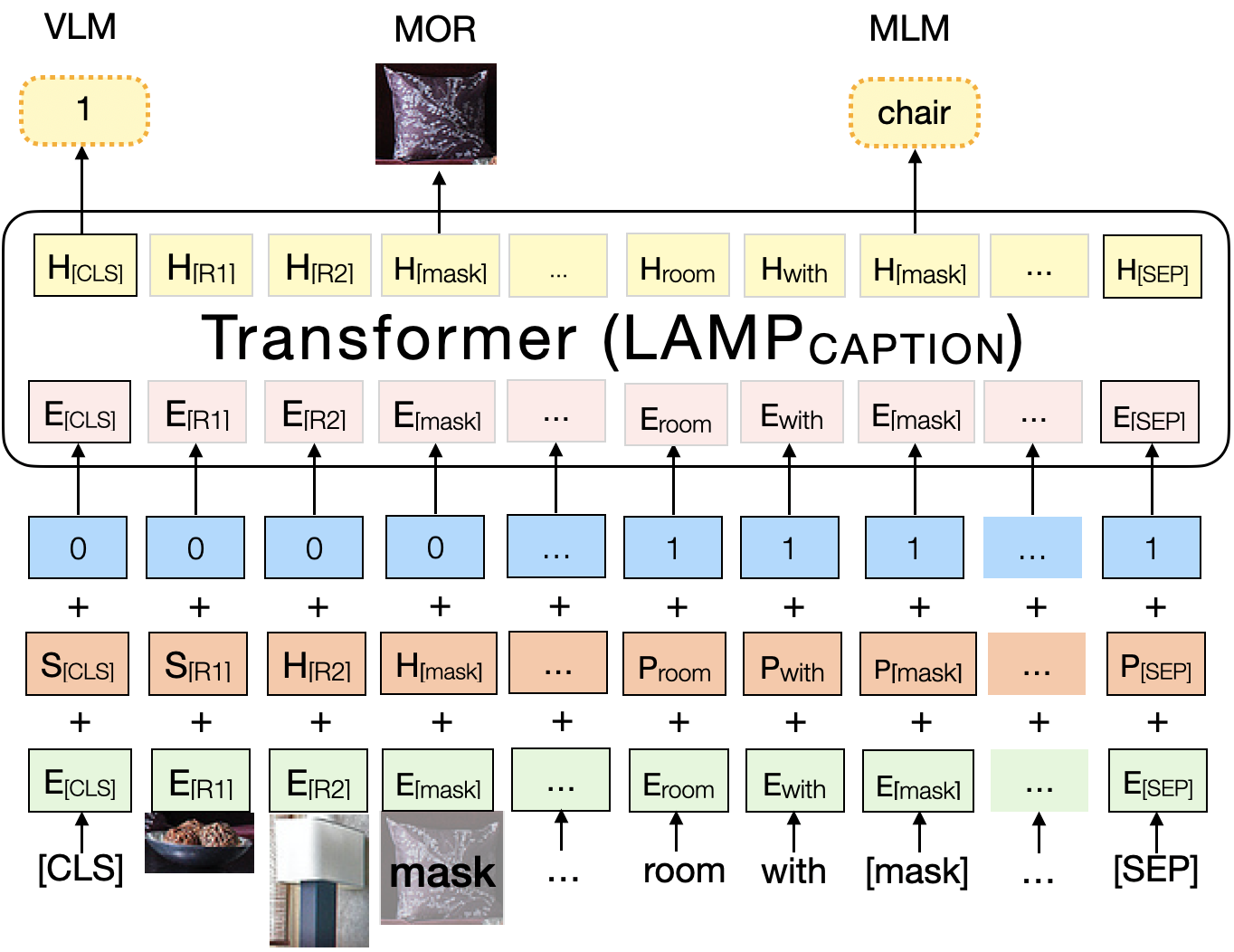}
\end{minipage}
}
\vspace{-0.3cm}
\caption{Overview of the proposed LAMP model. Both image-caption and image-labels are used in pretraining, to provide both fine-grained and coarse description to images. }
\vspace{-0.5cm}
\end{figure}

To address the above problems, we proposed a \textbf{L}abel-\textbf{A}ugmented \textbf{M}ultimodal \textbf{P}retraining model, named LAMP, where we leverage auto-generated labels of visual objects for better cross-modal alignment.
Specifically, our pretrained model follows the basic architecture of BERT. But different from the previous works, we extract objects-labels pairs by Faster-RCNN to propose a novel objects-entities level alignment pretraining task. So that when pretraining, all of visual objects have the corresponding fine-aligned tokens in lingual side. Besides, we also propose a second-stage pretraining, which serves as a bridge to reduce the discrepancy between pretraining and downstream dataset. The objects-labels pairs are also used to construct the fine-grained alignment.
In general, We proposed a two-stage pretraining method augmented by label textual information. Our contributions can be summarized as:

1. We presented a new V-L pretraining model, named LAMP, where a new object-label alignment task was designed to reduce the distortion, which is caused by the insufficient coverage of textual description for the corresponding image.

2. We proposed a label-augmented second-stage pretraining to reduce the discrepancy between pretraining and downstream tasks.

3. Our proposed model were evaluated in various experiment settings and achieved outstanding performances. Besides, we also conducted a series of ablation experiments to prove our motivations.







\section{Related Work}

In recent years, pretraining models have made great progress in both computer vision (CV) and natural language processing (NLP) communities. Generally, the design of pretraining paradigm is consisted of two parts: the design of pretraining tasks, and the design of model architecture. 

In vision community, most of pretraining models are based on CNN structure and use ImageNet Classification~\cite{deng2009imagenet} as the pretraining task. AlexNet~\cite{krizhevsky2012imagenet} is the pioneer in this field. After that, a series of architectures such as VGGNet~\cite{simonyan2014very} and ResNet~\cite{he2016deep} were proposed and achieved impressive results. These models are often used for feature extraction and have proven their values in various downstream tasks, such as object detection~\cite{girshick2014rich}, instance segmentation~\cite{hariharan2014simultaneous}, semantic segmentation~\cite{long2015fully}.


As for NLP community, most of the pretraining models are based on multi-layer Transformer, such as BERT~\cite{devlin2018bert}, GPT~\cite{gpt}, XLNet~\cite{yang2019xlnet} and RoBERTa~\cite{liu2019roberta}. Among them, BERT is perhaps the most classic and popular one due to its simplicity and outstanding performance. 
As a pioneering work, BERT proved that self-supervised learning, such as Masked Language Model task, is an effective way for universal representation learning in NLP.


More recently, there has been a surging interest in multi-modal representation learning, by pretraining on large-scale image/video and text pairs. VideoBERT~\cite{sun2019videobert} is the first work to perform BERT-based visual-linguistic pretraining. After VideoBERT, some researchers followed the approach of BERT and proposed some creative multi-modal pretaining model architectures. For example, ViLBert~\cite{lu2019vilbert} and LXMERT~\cite{tan2019lxmert} proposed a specifically designed two-stream architecture.
While VisualBERT~\cite{li2019visualbert}, UNICODER~\cite{li2019unicoder}, VL-BERT~\cite{su2019vl} and ImageBERT~\cite{qi2020imagebert} used a single stream architecture in contrary. 

Meanwhile, many powerful tasks for V-L pretraining, including instance level alignment and objects-entities level alignment, were also proposed to bridge the semantic gaps between image and text with image-caption dataset. 
For example, 
Visual-linguistic Matching~(VLM)~\cite{li2019unicoder, qi2020imagebert}, an instance level alignment task, is to predict whether an image and a text are matched.
Masked Language Modeling~(MLM)~\cite{sun2019videobert, li2019visualbert}, adapted from BERT, is to predict the category of a masked text token, given the information of visual side.
Similarly, Masked Vision Modeling~(MVM)~\cite{lu2019vilbert, tan2019lxmert} is to predict visual objects by the clues of textual side. The common MVM tasks include Masked Object Classification (MOC), Masked Object Classification with KL-Divergence (MOC-kl), and Masked Objects Regression (MOR). Although their objective functions are different, the intuitions behind them are similar to MLM, which is to predict the masked elements by the clues from other modalities. However, such tasks ignore the fact that the amount of information contained in the two modalities is different. Generally, a caption cannot provide comprehensive description for an image. And we believe this makes MVM pre-training tasks more difficult than MLM. To solve this problem, we propose to leverage the auto-generated labels of visual objects to improve the comprehensiveness of the linguistic information. 

\section{Approach}

In this section, we will first explain two motivations of our LAMP model~(i.e. label-augmented MVM task and label-augmented second-stage pretraining). Then we will introduce its details and the pretraining process.

We will use objects-caption to denote the input data used in pretraining tasks. Similarly, objects-sentence and objects-labels refer to the data of downstream,  and our auto-genereted objects-labels dataset respectively in the same way.



\subsection{MVM Task Augmented by Label}







Multi-modal pretraining is dedicated to construct alignment between two modalities. Such alignment exists in two levels, instance level and objects-entities (fine-grained) level. The latter alignment is achieved by a masked technique, first proposed by BERT, that is applied on vision and text respectively. But when coming into multi-modal field, such alignment expects that every masked object or entity has its counterpart in the other modality. And this is overlooked in the previous practice.

Depending on the modality of the forecast target, objects-entities level alignment in V-L field can be divided into two categories, namely masked language modeling~(MLM), predicting textual elements by visual side, and masked visual modeling~(MVM), predicting visual elements by the textual side.

A key assumption to guarantee the effectiveness of these approaches is that the information in one modality is enough to predict information in the other modality. But such assumption is not true. As shown in Figure~\ref{fig:example}, even by human beings, it is almost impossible to predict some visual objects~(i.e. pillow, plate) just by the caption. Thus generally, MVM is a harder task due to the insufficiency of textual information.



At first glance, we can solve this problem by collecting more captions for each image, which depict every visual objects in it. But this hardly works in practice. One reason is due to the labour cost, and the other is related to the model's representative power. For example, Visual Genome have nearly 50 captions with the above mentioned picture. But current pretraining paradigm won't allow us to encode such numbers of sentences in pretraining models, especially when these sentences barely have context relationships among each other.

We notice that labels contains rich semantic information of a given image, and shares the same position information with its corresponding objects, which will be beneficial to object-entity alignment.

Thus, We propose to automatically extract labels of visual objects to augment textual information in MVM task. And this process add no extra computation burden, since the previous MVM practice already use labels as supervised target. Another advantage of this one-to-one correspondence between labels and objects is that label texts can use the same position embedding as objects.





\textbf{Implement Details.} Given an image, we use auto-generated labels to construct objects-labels pairs. Labels and objects are generated with a Faster-RCNN pre-trained on Visual Genome dataset\cite{krishna2017visual}, which has a total of 1600 category labels and 400 attribute labels. We keep a constant number $N$ objects given one image.
For each visual object, we can get two labels, a category label and an attribute label. So we get $2N$ labels with each image.  We tokenize these labels, Then these labels and objects are input into LAMP together to apply the MOC and MOR tasks.
In terms of embedding, all these labels use the same word embedding mechanism as captions, and the same spatial embedding of the corresponding visual objects. Please note that during pretraining, we use objects-labels pairs to do MOC and MOR tasks. While objects-caption pairs are used to do MOR, VLM and MLM tasks. 



\subsection{Label-augmented Second-stage Pretraining}

The discrepancy between pretraining and finetuning is always a problem for pretrain-then-finetune learning. In V-L field, the problem becomes even more serious. For example, in VQA dataset ``what'' and ``how'' are very common words but both of them are rare in image-caption pairs, which are widely used for pretraining. 

Fortunately, second stage pretraining has been proved to be helpful to alleviate this problem in NLP. By a second stage pretraining, the representation knowledge pretrained earlier can adapt to the new distribution of downstream dataset. \cite{chronopoulou2019embarrassingly}
But when adapted in V-L representation learning, this approach need to be modified. Because image-caption pairs are required for MVM training, and questions are not designed on describing the images precisely.


To tackle this problem, we can still rely on the auto-generated objects-labels pairs to enrich the information in the textual side again. We use both objects-labels and objects-sentence pairs in second stage pretraining.

\textbf{Implement Details.} 
The second stage of pretraining is similar to the first stage, that involves training with both objects-sentence pairs and objects-labels paris. On some tasks such as VQA, training directly on objects-sentence pairs can be better than directly finetuning, while on other tasks such as NLVR2, it will fail. And jointly training on objects-sentence pairs and objects-labels paris will always bring better performance.
Due to the unclear correspondence between the text and image of some tasks, we remove VLM pretraining for these tasks. For example, we use VLM in visual question answering task, but not in NLVR2.

\subsection{LAMP}
In this section, we will first review the architecture of our LAMP. Then we introduce its input and pretraining tasks, respectively.

In LAMP, we use BERT-base as our base architecture. BERT is a multi-layer transformer based model proposed for NLP pretraining. With tokenized sub-words as input, BERT is pretrained with MLM(Masked Language Modeling and NSP(Next Sentence Prediction) tasks. For multi-modal learning, LAMP modified BERT from the aspect of input and pretraining tasks.

The input of LAMP consists of visual part and text part. Among them, the text part is also tokenized by the bert tokenizer, while the visual part uses a Faster-RCNN to extract features as input. The extracted features contains three parts, visual features, position information and labels. For a given image, we only keep 36 regions due to the limited computation resources. 

\subsubsection{Input Embedding}
Both visual and textual information are represented as vectors, and then be fed into the model.

\textbf{Text Token Embedding.} 
Following the practice of BERT, we first tokenized sentences as text input and embed these text tokens by vectors.
And we follow the idea in BERT to use position embedding to embed the position information.
Besides we use segment embedding to denote the modality of input, where zero represents text and one represents visual.
The final representation for each token is the summation of word embedding, position embedding and segment embedding.

\textbf{Visual Embedding.}
For each input image, we use a Faster-RCNN to extract a constant number of region features, and their corresponding positions.  The position are represented by a 7-d vector, including the bottom-left and top-right corner and height, width of image, and the fraction of image area covered respectively. Then the position vector and the region feature are projected to the same embedding space respectively. We also set segment embedding as mentioned above. Finally, the overall embedding are the summation of feature embedding, position embedding, and segment embedding. 

\textbf{Label Embedding.} For label embedding, we use the same word embedding and segment embedding as them in Text Token Embedding. Since these labels are not natural language, it is not appropriate to embed the positions with the setting of Text Token Embedding. Thus we use the spatial embeddings of the corresponding regions.

\subsubsection{Pretraining Tasks}

In order to learn a good multi-modal representation, we use four pretraining tasks, namely VLM, MLM, MOC and MOR, to align image and text, which can be classified into two categories~(i.e. instance level alignment and object-entity level alignment). Among them, MOC and MOR are enhanced by the objects-labels pairs.

\textbf{Instance Level Alignment.} We selected VLM task to align images and captions. 

\emph{VLM.} We added an additional token~[CLS] in our samples to represent the overall representation of objects-caption pairs. Then we fed the output of [CLS] into a fully-connected layer to predict whether an image and a caption are matched. During pretraining, we sampled 15\% of objects-caption pairs as positive samples and generate the same number of negative samples by replacing the caption with another random-selected one.

\textbf{Object-entity Level Alignment.} Here we selected MLM and two MVM tasks~(i.e. MOC and MOR) to align visual objects and textual entities.

\emph{MLM.}
In MLM, we followed the practice in standard BERT, masking approximately 15\% of sub-words, and asking the model to predict their category given the remaining inputs. Masked sub-words are replaced by a special token [MASK] 80\% of the time, or replaced by a random word 10\% of the time, the rest of them stay unchanged.

\emph{MVM}
Similar to MLM, we randomly sample 15\% of the objects to be masked. 80\% of the masked object features is set to zero, 10\% are replaced by a random feature, and the remaining 10\% remains unchanged. When a object is masked, the model is required to predict its category or its input feature, depending on whether the task is MOC or MOR. We use cross entropy for MOC and MSE for MOR respectively.
Please note again, we discard MOC task for objects-caption pairs during pretraining, because the insufficient textual information problem we mentioned before.






\subsubsection{Two-stage Pretraining}
The pretraining of LAMP consists of two stages. 
The first stage is to pretrain on both objects-caption pairs and objects-labels pairs. 
We apply MLM, VLM, and MOR for objects-caption pairs, while apply MOR and MOC for objects-labels pairs.

The second stage is to pretrain on task-specific data. Given a downstream task, we generally follow the idea of the first stage to pretrain on the dataset and auto-generated objects-labels pairs of the given downstream task. According to the specific downstream task, VLM may not be used in second-stage pretraining.

In a word, we use label-augmentation in both the first and second stages to improve performance.

\begin{table*}[!t]
\small
\begin{spacing}{1.2}
\centering \begin{tabular}{lccccc}
\toprule
\multicolumn{1}{c}{Task}                         & Datasets      & Image Src.                      & \#Images & \#Text  & Metric          \\ \midrule
\multicolumn{1}{c}{\multirow{2}{*}{1 Pre-train}} & MS-COCO       & --                              & 123K     & 617K & --              \\
\multicolumn{1}{c}{}                             & Visual Genome & --                              & 108K     & 5.39M  & --              \\ 
2 VQA                                            & VQA 2.0       & COCO                            & 204K     & 1.1M    & VQA-score       \\
3 GQA                                            & GQA           & COCO                            & 85K      & 22M     & GQA-score       \\
4 NLVR                                           & NLVR          & Web Crawled                     & 214K     & 107K    & Accuracy        \\ 
5 ZS IR                                          & Flickr30K     & \multicolumn{1}{l}{Web Crawled} & 1K       & 5K      & Recall@1, 5, 10 \\ \bottomrule
\end{tabular}
\caption{Statistics on the datasets used for pretraining and downstream tasks}
\end{spacing}
\label{Table1}
\vspace{-6mm}
\end{table*}

\section{Experiment}

        

In this section, we will first introduce the pretraining data and procedure for pretraining our LAMP model. And we will show our evaluation results on four downstream tasks and compare them with seven state-of-the-art models. A total of two models were obtained which are trained with MSCOCO and MSCOCO+VG respectively. We will share our data and models~\footnote{\url{ https://docs.google.com/document/d/1FzZIXNO8e2Qj2xB2JOm0uXiwPC3DxWyUjzMQKARghWo/edit?usp=sharing}}.

\subsection{Pretrianing Data}
We use two datasets, MSCOCO~\cite{lin2014microsoft} and Visual Genome~(VG) datasets~\cite{krishna2017visual}, for pretraining our LAMP model.

We select train and dev sets of MSCOCO and VG for pretraining to avoid data leakage in downstream tasks. 5K data in MSCOCO dev sets is kept for validation.
And for each image, we selected 36 detected regions by their confidence values and used a public Faster-RCNN~\cite{ren2015faster, anderson2018bottom} to generate their labels.
Removing the overlapping part of MSCOCO and VG, The final pretraining dataset totally contains 175K images with 63M labels distributed in 2000 categories, and 5.9M image-caption pairs. 



\subsection{Pretraining Procedure}
As for the model architecture, we used a Transformer with 12 layers, where each layer had 768 hidden units and 12 self-attention heads. The maximum sequence length of text was set to 20. Parameters were initialized with BERT-base.
We trained the objects-caption pairs and objects-labels pairs all with batch size of 256 respectively. 

The first stage pretraining took 20 epochs while the second one took 10 epochs.
We used Adam~\cite{kingma2014adam} as the optimizer with initial learning rates of 1e-4. And a linear decay learning rate schedule with warm up was applied. Experiments were conducted on Titan RTXs and GTX 1080Tis, and all experiments can be replicated on at most 2 Titan RTXs each with 24GBs of GPU memory. We also used apex to accelerate training and save GPU memory. 

Pretraining on MSCOCO generally takes less than 10 hours on 2 Titan RTXs while task-speciﬁc pretraining and ﬁnetuning will take much less. 




\subsection{Fine-tuning on Downstream Tasks}

\begin{table*}[!t]
\small
\begin{spacing}{1.2}
\centering \begin{tabular}{clcclclclccc}
\toprule
 & \multicolumn{1}{c}{\multirow{2}{*}{Method}} & \multicolumn{2}{c}{VQA} &  & GQA  &  & NLVR     &                      & \multicolumn{3}{c}{ZS Image Retrieval} \\ \cline{3-4}
 & \multicolumn{1}{c}{}                        & test-dev   & test-std   &  & test &  & test-P & \multicolumn{1}{c}{} & R1          & R5         & R10         \\ \hline
\multirow{7}{*}{SOTA} & \multicolumn{1}{l}{ImageBERT(with LAIT)}  & \multicolumn{1}{c}{--} &  \multicolumn{1}{c}{--} & & \multicolumn{1}{c}{--} & & \multicolumn{1}{c}{--} & & \textbf{54.3} & \textbf{79.6} & \textbf{87.5} \\
& 
\multicolumn{1}{l}{ImageBERT(w/o LAIT)}  & \multicolumn{1}{c}{--} &  \multicolumn{1}{c}{--} & & \multicolumn{1}{c}{--} & & \multicolumn{1}{c}{--} & & 48.4 & 76.0 & 85.2 \\
                      & VisualBERT  & 70.80  & 71.01 &  & --    &  & 67.0  &  & --    & --   & --   \\
                      & VL-BERT$_{LARGE}$     & 71.79  & 72.22 &  & --    &  & --    &  & --    & --   & --   \\
                      & LXMERT$_{LARGE}$      & 72.42  & 72.54 &  & 60.30   &  & 74.5  &  & --    & --   & --   \\
                      & Unicoder-VL & --     & --    &  & --    &  & --    &  & 48.4  & 76.0 & 85.2 \\
                      & ViLBERT     & 70.55  & 70.92 &  & --    &  & --    &  & 31.9  & 61.1 & 72.8 \\ 
\multicolumn{2}{l}{LAMP~(MSCOCO)} & 70.85  & 71.0 &  & -- &  & 74.34  &  & 42.5 & 70.9 & 80.8 \\
\multicolumn{2}{l}{LAMP~(MSCOCO+VG)}  & \textbf{72.48}  & \textbf{72.62} &  & \textbf{61.05} &  & \textbf{75.43}  &  & 51.8 & 77.4 & 85.3\\ \bottomrule
\end{tabular}
\vspace{-2mm}
\caption{Results of downstream tasks. UNITER and ImageBERT are compared in the fair settings.
}
\label{Table2}
\end{spacing}
\vspace{-5mm}
\end{table*}

We finetuned our LAMP model on three downstream tasks by transferring the pre-trained model to each target task. To directly test the model and avoid the impact of different finetune strategies, we additionally conducted a zero-shot task. We will describe the problem, dataset, model modifications, and training objective for each task below. 



\textbf{Visual Question Answering (VQA).} The VQA task is, given a natural image, to select the correct answer for a perceptual-level question. We trained and evaluated on the VQA 2.0 dataset~\cite{antol2015vqa}, consisting 1.1M questions about MSCOCO images, each with 10 ground truth answers.

We treat the task of VQA as a classification problem. On VQA, we built a classifier upon the output representation of [CLS].
And before finetuning, we did the second-stage pretraining on VGQA, GQA and VQA datasets with MLM, MOC, MOR, VLM and visual question task for 10 epochs with 1-e4 learning rate. Label information is used of the same way as in first stage pretraining. Then the pretrained model was finetuned by 4 epochs with 5e-5 learning rate.



\textbf{GQA.}
The task of GQA~\cite{hudson2019gqa} is the same as VQA, but GQA requires stronger reasoning ability (e.g., spatial understanding and multi-step inference). GQA generated 22M questions about MSCOCO images from ground truth image scene graph for explicitly controlling question quality.

We used the same training procedure as that in VQA. After pretraining with MSCOCO and VG, we continued to use VGQA~(Visual Genome Question Answering), GQA, and VQA to do second-stage pretraining augmented by label information, and then finetuned on GQA.
To finetune LAMP on GQA, we also built a classifier upon the output representation of [CLS].

\textbf{NLVR2.}
Each datum in NLVR2~\cite{suhr2018corpus} contains two related natural images and one natural language statement. The dataset consists of over 100K examples of English sentences paired with web images. The task of NLVR2 is to predict whether the statement correctly describes these two images or not. Unlike VQA and GQA dataset, all the sentences and images of NLVR2 are not covered in pre-training data. 

We used both label information and NLVR2 data to do the second-stage pretraining. We found that in this setting, the second stage failed if we did not use label.
To finetune LAMP on NLVR2, we built a classifier upon the output representation of [CLS] of both sentences corresponding to a image. 

\textbf{Flickr30K.} This dataset contains 31,783 images collected from the Flickr website, each with five captions. We used this dataset to do the zero-shot image retrieval(ZS IR), in which we directly applied the pretrained VLM prediction mechanism to caption-based image retrieval without finetuning. Through this task, our model demonstrates excellent multi-modal representation capabilities without finetuning. Following ViL-BERT\cite{lu2019vilbert}, we chose 1000 images and 5000 corresponding captions to perform zero short tasks. We used three evaluation metrics, i.e., R@K (K=1,5,10), where R@K was the percentage of ground-truth matchings occurred in the top K-ranked results. 

\subsection{Results on Downstream Tasks}

In this section, we compare our method with some of the strong performance models proposed recently, LXMERT, VisualBERT, ViLBERT, VLBERT, UNITER, ImageBERT, and UNICODER-VL. 
These models are almost the state-of-the-art models in terms of model design, training task combination, and training dataset.
And although most of them leverage labels of visual objects as supervised signals, the alignment between the textual information of labels and the visual objects is overlooked.
ImageBERT is trained on Conceptual Caption, SBU caption, and an additional caption dataset containing 10M image-caption pairs, called LAIT. We report two results of zero shot image retrieval of ImageBERT trained with and without the LAIT.
UNITER has better results with additional pretraining dataset, and we used the results under comparable pretraining and finetuning settings for fairness.

The overall performances of our LAMP and baselines are shown in Table~\ref{Table2}. 
These downstream tasks can be divided into finetuning and zero shot settings. Next we will discuss them, respectively.

\textbf{Fine-tuning Tasks.}
We compare our model with other state-of-the-art BERT-based multi-modal pretrained models. The results in Table~\ref{Table2} demonstrate the strong performances of LAMP. In VQA and GQA, LAMP with MSCOCO+VG achieved the best performances. In NLVR2, our model got a comparable performance with UNITER, which has the best results right now. AND LAMP trained with MSCOCO only achieve remarkable result on NLVR2 compared with LXMERT and VisualBERT. 

\textbf{Zero Shot Image Retrieval Task.}
UNITER and ImageBERT achieves better results in this task. However, they use a much larger corpus in pretraining procedure. In specific, UNITER combines four datasets (Conceptual Captions, SBU Captions, VG and MSCOCO) together to generate a 9.6M training corpus. And ImageBERT aggregates a dataset consisted of more than 10M images. Both Unicoder-VL and ImageBERT use over features of 100 objects per image. Considering LAMP just use 36 objects, they demonstrate that more objects can lead to better result. 
Even so, we still outperform UNICODER.  LAMP(MSCOCO) also establishes a great baseline on this task, exceeding ViLBERT over 8\% with a relative small pretraining dataset.

\section{Analysis} 
In this section, we will first test the effectiveness of label augmentation in pretraining tasks. We will also evaluate the performance of second-stage pretraining on various experimental settings.

\subsection{Evaluation of Different Pretrained Models on MSCOCO}
The difficulty in comparing different pretraining models is that pretraining data and model size are different among them. This prevents us from having a insight of the different pretraining tasks and architecture.
We use MSCOCO as the benchmark dataset to compare our model with two representative models, LXMERT and VisualBERT, which are a single-stream model and a two-stream model respectively. In addition, they were all pretrained on MSCOCO and finetune on two tasks, VQA and NLVR2 respectively. 
The results are shown in Table~\ref{Table3}.
The released VisualBERT was just pretrained on MSCOCO so we directly used its published results. But the released LXMERT is also pretrained on much larger data~(i.e. VG, VQA, GQA, and VGQA). So we retrained LXMERT on MSCOCO with their own hyperparameters.

\begin{table}[h]
\footnotesize
\centering\begin{tabular}{lllllllll}
\toprule
\multirow{2}{*}{Method} &
   
  \multicolumn{1}{c}{VQA} &
   
  \multicolumn{1}{c}{NLVR} &
  \\
 & 
  \multicolumn{1}{c}{test-dev} &
   
  \multicolumn{1}{c}{test-P} &
 \\
\midrule
\multicolumn{1}{l}{LAMP}         & \multicolumn{1}{c}{70.05} &   \multicolumn{1}{c}{74.34} &   \\
\multicolumn{1}{l}{LAMP with 2nd stage pretraining (VQA)}        & \multicolumn{1}{c}{70.85} &   \multicolumn{1}{c}{--} &    \\
\multicolumn{1}{l}{VisualBERT}              & \multicolumn{1}{c}{70.8} &   \multicolumn{1}{c}{67.0} &   \\
\multicolumn{1}{l}{LXMERT (MSCOCO)}         & \multicolumn{1}{c}{68.62} & \multicolumn{1}{c}{66.79} &   \\
\multicolumn{1}{l}{LXMERT (Full Version)}                  & \multicolumn{1}{c}{72.42} & \multicolumn{1}{c}{74.5} &   \\
\bottomrule
\end{tabular}
\caption{Comparison of pretraining models pretrained on MSCOCO dataset.}
\label{Table3}
\vspace{-3mm}
\end{table}

We can see with this same setting, LAMP achieved impressive result.  VisualBERT also used a second-stage pretraining and got good results on VQA, but it was not as good as us on NLVR2. There is a big gap between the full training version of LXMERT and the MSCOCO version. We speculate that it is related to the size of the dataset and to the specific visual answering task of LXMERT.
It is worth noting that on the NLVR2 task,  LAMP with MSCOCO has achieved similar result compared with the full training version of LXMERT.

\subsection{Evaluation of Label Augmentation on Pretraining Tasks}

To validate the effectiveness of label augmentation in pretraining, We conducted extensive experiments with different combinations of losses.

The results are shown in Table~\ref{Table4}. We can find that the label augmentation has a stable improvement in all of situations, while MOC generally harms the performance in downstream tasks.
\begin{table}[h]
\small
\centering\begin{tabular}{llllllll}
\toprule
\multirow{2}{*}{Method} &
   
  \multicolumn{1}{c}{VQA} &
   
  \multicolumn{1}{c}{NLVR} 
  \\
 & 
  \multicolumn{1}{c}{dev} &
  \multicolumn{1}{c}{dev} 
 \\
\midrule
\multicolumn{1}{l}{LM + VLM}                      & \multicolumn{1}{c}{66.80} &                 \multicolumn{1}{c}{73.35}    \\
\multicolumn{1}{l}{LM + VLM + MOR}                 & \multicolumn{1}{c}{66.91} &   \multicolumn{1}{c}{73.89}   \\
\multicolumn{1}{l}{LM + VLM + MOR + MOC}           & \multicolumn{1}{c}{65.57} &   \multicolumn{1}{c}{71.40}    \\
\multicolumn{1}{l}{LM + VLM + Label}               &  \multicolumn{1}{c}{67.23}&   \multicolumn{1}{c}{73.70}   \\
\multicolumn{1}{l}{\textbf{LM + VLM + MOR + Label (ours)}}         & \multicolumn{1}{c}{\textbf{67.50}} &   \multicolumn{1}{c}{\textbf{74.42}}    \\
\multicolumn{1}{l}{LM + VLM + MOR + MOC + Label}   & \multicolumn{1}{c}{67.49} &   \multicolumn{1}{c}{73.35}     \\
\bottomrule
\end{tabular}
\vspace{-1mm}
\caption{Evaluation of pretrained model with different pretraining tasks.}
\label{Table4}
\vspace{-4mm}
\end{table}

\subsection{Evaluation of Label-Augmented Second-stage Pretraining}
In this section, we test the effectiveness of label-augmented second-stage pretraining on VQA and NLVR2. The results are shown in Table~\ref{Table5}.

\textbf{VQA.}
In VQA task, three different strategies are applied on the first stage LAMP pretrained on MSCOCO+VG. We first directly finetune on VQA dataset. The result is already better than several models (VLBERT, VisualBERT, and VilBERT) with this setting. Then we apply the second stage pretraining with and without label augmentation, achieving 71.77 and 71.9 on VQA test-std, without any other data augmentation. The second stage pretraining with label augmentation effectively improves the performance of VQA.

\textbf{NLVR2.}
Unlike VQA dataset, a sentence in NLVR2 describes the relationship of two pictures, which make the label augmentation much more important. We only use MLM and MOR with NLVR2 dataset, because the task of VLM conflicts with NLVR2 task.
Specifically, we conducted three experiments. First we directly finetuned on NLVR2 task, achieving 75.74 accuracy. Then we conducted second stage pretraining with and without label augmentation. It turns out that only with label dataset can make second stage pretraining useful. 


\begin{table}[h]
\footnotesize
\centering\begin{tabular}{lccc}
\toprule
\multirow{2}{*}{Method} & \multicolumn{2}{c}{VQA} & NVLR2 \\
 & test\_dev & test\_std & dev \\ \midrule
Without 2nd stage pretraining & 71.35 & 71.53 & 75.74 \\
2nd stage MLM pretraining & 71.74 & 71.77 & 75.12 \\
2nd stage pretraining with VQA task & 71.93 & 71.9 & 76.41 \\ \bottomrule
\end{tabular}
\caption{Evaluation of second stage pretraining on VQA and NLVR tasks (MSCOCO + VG).}
\label{Table5}
\vspace{-5mm}
\end{table}

\section{Conclusion}
In conclusion, to solve the insufficiency of textual information and the discrepancy between pretraining tasks and downstream tasks, we leveraged auto-generated labels of visual objects and proposed a label-augmented pretraining model, named LAMP. The extensive experiments well proved the value of labels in V-L pretraining and the effectiveness of LAMP.


{\small
\bibliographystyle{ieee_fullname}
\bibliography{egbib}
}
\end{document}